# Two-vector representation of a nondepolarizing Mueller matrix

José J. Gil[1*], Ignacio San José[2]

[1]Universidad de Zaragoza. Pedro Cerbuna 12, 50009 Zaragoza, Spain.
[2]Instituto Aragonés de Estadística. Gobierno de Aragón. Bernardino Ramazzini 5, 50015 Zaragoza, Spain.
*Corresponding author: ppgil@unizar.es

**Abstract**

A geometric view of the polarimetric properties of a nondepolarizing medium is presented by means of a pair of vectors in the Poincaré sphere. An alternative representation constituted by a set of vectors contained in the equatorial plane of the Poincaré sphere is also defined and interpreted. The analyses of the magnitudes and relative orientations of the constitutive vectors of such simple representations allow for a classification of nondepolarizing media.

## 1. Introduction

Nondepolarizing (or *pure*) Mueller matrices constitute a key subset of Mueller matrices because 1) they represent the basic linear polarimetric interactions, so that any general Mueller matrix can be considered as an ensemble average of pure Mueller matrices [1-4], and 2) they, by themselves, represent a wide range of polarimetric behaviors of media, frequently encountered as a good estimate of experimentally measured Mueller matrices [5,6]. Therefore, their interpretation and geometric representation through simple and meaningful structures deserve particular attention.

In this work we introduce a representation of the normalized Mueller matrix $\hat{\mathbf{M}}_J$ of a nondepolarizing medium by means of a pair of vectors in the Poincaré sphere, whose magnitudes and relative orientations provide direct information of the main features of $\hat{\mathbf{M}}_J$. An alternative representation, contained in the equatorial plane of the Poincaré sphere, is also presented on the basis of four two-dimensional vectors, two of which being horizontal.

As in other related works, and for the sake of simplicity of the mathematical formulation of some relations, we will take advantage of the partitioned block expression of a Mueller matrix [7],

$$\mathbf{M} = m_{00}\hat{\mathbf{M}}$$

$$\hat{\mathbf{M}} \equiv \begin{pmatrix} 1 & \mathbf{D}^T \\ \mathbf{P} & \mathbf{m} \end{pmatrix} \quad \mathbf{m} \equiv \frac{1}{m_{00}}\begin{pmatrix} m_{11} & m_{12} & m_{13} \\ m_{21} & m_{22} & m_{23} \\ m_{31} & m_{32} & m_{33} \end{pmatrix}$$

$$\mathbf{D} \equiv \frac{1}{m_{00}}(m_{01}, m_{02}, m_{03})^T \quad \mathbf{P} \equiv \frac{1}{m_{00}}(m_{10}, m_{20}, m_{30})^T \quad (1)$$

where $\hat{\mathbf{M}}$ is the *normalized Mueller matrix* and $m_{00}$ is the *mean intensity coefficient* (i.e., the *transmittance*, or *gain* of $\mathbf{M}$ for input unpolarized light), while the $\mathbf{D}$ and $\mathbf{P}$ are respectively called the *diattenuation vector* and the *polarizance vector* of $\mathbf{M}$ [8], whose absolute values are the *diattenuation*, $D \equiv |\mathbf{D}|$ and the *polarizance*, $P \equiv |\mathbf{P}|$ of $\mathbf{M}$.

It is worth recalling that both polarizance $P$ and diattenuation $D$ share a common physical nature [9,10] because $D$ is both the diattenuation of $\mathbf{M}$ and the polarizance of the *reverse Mueller matrix* [11,12]

$$\mathbf{M}^r \equiv \text{diag}(1,1,-1,1)\mathbf{M}^T \text{diag}(1,1,-1,1) \quad (2)$$

corresponding to the same interaction as $\mathbf{M}$, but interchanging the input and output directions ($\mathbf{M}^T$ being the transposed matrix of $\mathbf{M}$). Conversely, $P$ is both the polarizance of $\mathbf{M}$ and the diattenuation of $\mathbf{M}^r$. Moreover, the *degree of spherical purity*, defined as

$$P_s = \|\mathbf{m}\|_2/\sqrt{3} \quad (3)$$

where $\|\mathbf{m}\|_2$ stands for the Frobenius norm of the 3x3 submatrix $\mathbf{m}$, gives a measure of how close is $\mathbf{M}$ to the Mueller matrix of a retarder [10,13] (irrespectively of the value of the retardance).

A given Mueller matrix $\mathbf{M}$ is said to be *pure* (or *nondepolarizing*) when it transforms any input totally polarized state into a corresponding output totally polarized state, that is, when the *depolarization index* $P_\Delta$ (also called *degree of polarimetric purity* of $\mathbf{M}$) [14], satisfies $P_\Delta = 1$, i.e.,

$$P_\Delta^2 = 1 = (2D^2 + 3P_s^2)/3 \quad (4)$$

where the equality $D = P$ that holds for any pure Mueller matrix (16) has been taken into account. Note that the term *pure* used to refer to Mueller matrices satisfying $P_\Delta = 1$ comes from their statistical nature, that is, as indicated above, any nonpure Mueller matrix can always be expressed as a convex combination (or an ensemble average) of pure Mueller matrices. Hereafter, as usual, we will use the common notation $\mathbf{M}_J$ to distinguish pure Mueller matrices from general Mueller matrices ($\mathbf{M}$).

From a practical point of view, it is worth observing that, due to the limited precision of the polarimetric setups, experimentally determined Mueller matrices are affected by errors and it is necessary to use appropriate criteria for deciding if a measured Mueller matrix $\mathbf{M}_{\exp}$ is considered pure in practice. In fact, when the value of $|1 - P_\Delta| \approx \delta$, where $\delta$ is the predetermined tolerance, $\mathbf{M}_{\exp}$ can be submitted to a filtering process [5,15], in such a manner that the filtered matrix $\mathbf{M}_f$ is a Mueller matrix that satisfies $P_\Delta(\mathbf{M}_f) = 1$, that is, $\mathbf{M}_f$ satisfies exactly the following characteristic property of pure Mueller matrices [7]

$$\mathbf{G}\mathbf{M}_f^T\mathbf{G}\mathbf{M}_f = \sqrt{\det \mathbf{M}_f}\, \mathbf{I}_4$$

$$[\mathbf{G} \equiv \text{diag}(1,-1,-1,-1) \quad \mathbf{I}_4 \equiv \text{diag}(1,1,1,1)] \quad (5)$$

## 2. Block expression for the polar decomposition of a pure Mueller matrix

It is well-known that any pure Mueller matrix $\mathbf{M}_J$ can always be expressed as the product of the Mueller matrices of a diattenuator (in general elliptic) and a retarder (in general elliptic) [17,18,8] in either of the two possible relative positions

$$\mathbf{M}_J = \mathbf{M}_R \mathbf{M}_D = \mathbf{M}_P \mathbf{M}_R \quad (6)$$

where $\mathbf{M}_R$ represents a retarder, while the right-equivalent and left-equivalent diattenuators, $\mathbf{M}_D$ and $\mathbf{M}_P$, are mutually related through

$$\mathbf{M}_P = \mathbf{M}_R \mathbf{M}_D \mathbf{M}_R^T \quad (7)$$

In order to express Eq. (6) in terms of the block parts of $\mathbf{M}_J$, let us recall the general block form of the pure Mueller matrix of a retarder

$$\mathbf{M}_R \equiv \begin{pmatrix} 1 & \mathbf{0}^T \\ \mathbf{0} & \mathbf{m}_R \end{pmatrix} \quad \left( \mathbf{m}_R^{-1} = \mathbf{m}_R^T \quad \det \mathbf{m}_R = +1 \right) \quad (8)$$

as well as the block form of the pure Mueller matrix of a normal diattenuator

$$\mathbf{M}_D \equiv m_{00} \begin{pmatrix} 1 & \mathbf{D}^T \\ \mathbf{D} & \mathbf{m}_D \end{pmatrix}$$

$$\mathbf{m}_D \equiv \sin\kappa \, \mathbf{I}_3 + (1-\sin\kappa) \hat{\mathbf{D}} \otimes \hat{\mathbf{D}}^T$$

$$\left( \sin\kappa \equiv \sqrt{1-D^2} \quad \mathbf{I}_3 \equiv \mathrm{diag}(1,1,1) \quad \hat{\mathbf{D}} \equiv \mathbf{D}/D \right) \quad (9)$$

where $\sin\kappa$ (with $0 \le \kappa \le \pi/2$) is the *counterpolarizance* (or *counterdiattenuation*) of $\mathbf{M}_D$.

Therefore, Eq. (6) can be expressed as

$$\mathbf{M}_J = m_{00} \begin{pmatrix} 1 & \mathbf{D}^T \\ \mathbf{m}_R \mathbf{D} & \mathbf{m}_R \mathbf{m}_D \end{pmatrix} = m_{00} \begin{pmatrix} 1 & \left(\mathbf{m}_R^T \mathbf{P}\right)^T \\ \mathbf{P} & \mathbf{m}_P \mathbf{m}_R \end{pmatrix} \quad (10)$$

so that the following relations necessarily hold

$$\mathbf{P} = \mathbf{m}_R \mathbf{D} \quad \mathbf{D} = \mathbf{m}_R^T \mathbf{P}$$

$$\mathbf{m} = \mathbf{m}_R \mathbf{m}_D = \mathbf{m}_P \mathbf{m}_R \quad (11)$$

Note that $\hat{\mathbf{M}}_D$ and $\hat{\mathbf{M}}_P$ are fully determined by $\mathbf{D}$ and $\mathbf{P}$ respectively [see Eq. (9)] (with $D = P$). The action of $\mathbf{M}_R$ is characterized through its corresponding pair of eigenstates together with its retardance $\Delta$ (phase shift introduced between them). The eigenstates of $\mathbf{M}_R$ can be represented by the totally polarized Stokes vectors

$$\mathbf{s}_{R+} \equiv \begin{pmatrix} 1 \\ \mathbf{u}_R \end{pmatrix} \quad \mathbf{s}_{R-} \equiv \begin{pmatrix} 1 \\ -\mathbf{u}_R \end{pmatrix} \quad (12)$$

where the unit vector

$$\mathbf{u}_R \equiv \begin{pmatrix} \cos 2\varphi \cos 2\chi \\ \sin 2\varphi \cos 2\chi \\ \sin 2\chi \end{pmatrix} \quad (13)$$

is the *Poincaré vector* of the fast eigenstate of $\mathbf{M}_R$, with azimuth $\varphi$ and ellipticity angle $\chi$, while $(\varphi + \pi/2, -\chi)$ are the azimuth and ellipticity angle of the slow eigenstate, whose Poincaré vector $-\mathbf{u}_R$ is given by the antipodal point of $\mathbf{u}_R$ on the Poincaré sphere.

Consequently, both $\mathbf{m}_R$ and $\mathbf{M}_R$ are fully determined by the corresponding *straight retardance vector*, defined as

$$\bar{\mathbf{R}} \equiv \frac{\Delta}{\pi} \mathbf{u}_R \quad (14)$$

Observe that, unlike the retardance vector $\mathbf{R} \equiv \Delta \mathbf{u}_R$ [8], $\bar{\mathbf{R}}$ has been defined in such a manner that its absolute value $\bar{R} \equiv |\bar{\mathbf{R}}|$ is restricted to the range $0 \le \bar{R} \le 1$, and thus, as occurs with $\mathbf{D}$ and $\mathbf{P}$, $\bar{\mathbf{R}}$ can be represented by a point in the Poincaré sphere. The maximum retardance $\Delta = \pi$ (with the usual convention $0 \le \Delta \le \pi$) corresponds to $\bar{R} = 1$, in which case the end point of $\bar{\mathbf{R}}$ lies on the surface of the Poincaré sphere, while $\Delta = 0$ corresponds to $\bar{R} = 0$, in which case the end point of $\bar{\mathbf{R}}$ coincides with the origin of the Poincaré sphere.

In summary, the *constitutive vectors* $\mathbf{D}$, $\mathbf{P}$ and $\bar{\mathbf{R}}$ of a nondepolarizing Mueller matrix $\mathbf{M}_J$ contain complete information about $\hat{\mathbf{M}}_J$, that can be represented by means of the respective points in the Poincaré sphere. Nevertheless, since $\hat{\mathbf{M}}_J$ depends on up to six real parameters, the constitutive vectors are not mutually independent and a *two-vector representation* of $\hat{\mathbf{M}}_J$ is dealt with in Section 5.

## 3. Block expression for the dual linear retarder decomposition of a pure Mueller matrix

Ossikovski showed that any pure Mueller matrix can be expressed as [19]

$$\mathbf{M}_J = m_{00} \mathbf{M}_{RL2}(\varphi_2, \Delta_2) \hat{\mathbf{M}}_{RDL0}(D, \Delta_0) \mathbf{M}_{RL1}(\varphi_1, \Delta_1) \quad (15)$$

where $\mathbf{M}_{RL1}$ and $\mathbf{M}_{RL2}$ are the Mueller matrices of respective *entrance* and *exit* linear retarders, with respective azimuths $(\varphi_1, \varphi_2)$ and retardances $(\Delta_1, \Delta_2)$, while the central matrix $\mathbf{M}_{RDL0}$ represents a linear *diattenuating retarder* oriented at 0º, with diattenuation $D$ and retardance $\Delta_0$, that is, $\mathbf{M}_{RDL0}$ corresponds to a serial combination of a linear retarder $\mathbf{M}_{RL0}(\Delta_0)$ and a normalized linear diattenuator $\mathbf{M}_{DL0}(D)$, both with the same orientation at 0º with respect to the reference axes

$\mathbf{M}_{RDL0}$ is a normal matrix, with two mutually orthogonal linear eigenstates

$$\mathbf{s}_{1+} \equiv (1,1,0,0)^T \quad \mathbf{s}_{1-} \equiv (1,-1,0,0)^T \quad (16)$$

which coincide with the eigenstates of the said serial components.

Consequently $\mathbf{M}_{RL0}$ and $\mathbf{M}_{DL0}$ commute [19],

$$\mathbf{M}_{RDL0} = \mathbf{M}_{RL0}(\Delta_0) \mathbf{M}_{DL0}(D)$$
$$= \mathbf{M}_{DL0}(D) \mathbf{M}_{RL0}(\Delta_0) \quad (17)$$

so that $\hat{\mathbf{M}}_{RDL0}(D, \Delta_0)$ has the block form

$$\mathbf{M}_{RDL0} = m_{00} \begin{pmatrix} 1 & \mathbf{D}_0^T \\ \mathbf{D}_0 & \mathbf{m}_{RDL0} \end{pmatrix} \quad (18.a)$$

with

$$\mathbf{D}_0 \equiv (\cos\kappa, 0, 0)^T$$

$$\mathbf{m}_{RDL0} \equiv \begin{pmatrix} 1 & 0 & 0 \\ 0 & \sin\kappa \cos\Delta_0 & \sin\kappa \sin\Delta_0 \\ 0 & -\sin\kappa \sin\Delta_0 & \sin\kappa \cos\Delta_0 \end{pmatrix}$$

$$(D \equiv \cos\kappa \quad \sqrt{1-D^2} \equiv \sin\kappa \quad 0 \le \kappa \le \pi/2) \quad (18.b)$$

Note that $\mathbf{M}_{DL0}$ can also be expressed as the following serial combination

$$\hat{\mathbf{M}}_{RDL0}(D,\Delta_0) = \mathbf{M}_{RL0}\left(\frac{\Delta_0}{2}\right)\hat{\mathbf{M}}_{DL0}(D)\,\mathbf{M}_{RL0}\left(\frac{\Delta_0}{2}\right) \quad (19)$$

where the retardance $\Delta_0$ has been decoupled from the *intrinsic diattenuator* $\hat{\mathbf{M}}_{DL0}(D)$ by splitting $\mathbf{M}_{RL0}(\Delta_0)$ into two identical linear horizontal retarders that sandwich $\hat{\mathbf{M}}_{DL0}(D)$. Therefore

$$\mathbf{M}_J = m_{00}\,\mathbf{M}_{RL2}\,\mathbf{M}_{RL0}\left(\frac{\Delta_0}{2}\right)\hat{\mathbf{M}}_{DL0}(D)\,\mathbf{M}_{RL0}\left(\frac{\Delta_0}{2}\right)\mathbf{M}_{RL1} \quad (20)$$

From Eqs. (**15**) and (**18**), $\mathbf{M}_J$ can be expressed in the following block form

$$\mathbf{M}_J \equiv m_{00}\begin{pmatrix} 1 & \left(\mathbf{m}_{RL1}^T\mathbf{D}_0\right)^T \\ \mathbf{m}_{RL2}\mathbf{D}_0 & \mathbf{m}_{RL2}\mathbf{m}_{RDL0}\mathbf{m}_{RL1} \end{pmatrix} \quad (21)$$

Observe that, since the diattenuation $D$ is not affected by the multiplication of $\mathbf{M}_{RDL0}$ by retarders, then $D(\mathbf{M}_J) = D(\mathbf{M}_{RDL0})$. As shown in Section 6, the fact that the dual linear retarder decomposition (**15**), with block expression (**21**), only contains serial components with respective linear eigenstates, allows for an interesting representation of $\hat{\mathbf{M}}_J$ by means of vectors lying in the equatorial plane of the Poincaré sphere.

## 4. Normality and degeneracy of nondepolarizing Mueller matrices

The notions of *normality* [20,21] (or homogeneity [22]) and *degeneracy* of nondepolarizing systems, developed in previous works under the scope of Jones algebra [22,23] are particularly useful for the analysis and understanding of the geometric representation of a pure Mueller matrix $\mathbf{M}_J$ by means of its constitutive vectors $\mathbf{D}, \mathbf{P}$ and $\bar{\mathbf{R}}$.

$\mathbf{M}_J$ is said to be *normal* (or *homogeneous*) when it satisfies the commutation property $\mathbf{M}_J\mathbf{M}_J^T = \mathbf{M}_J^T\mathbf{M}_J$ (which corresponds to the conventional notion of normality in matrix algebra). Normal pure Mueller matrices are characterized by the fact that they have two orthogonal pure eigenstates (represented by two totally polarized Stokes vectors that correspond to antipodal points on the Poincaré sphere). When $\mathbf{M}_J$ has two eigenstates that are not orthogonal, it is called *nonnormal* (or *inhomogeneous*). When the eigenstates degenerate into an only eigenstate (that is, both eigenstates coincide), its eigenvalue is doubly degenerate [22] and the corresponding Jones matrix is called *degenerate*.

Since Jones and pure Mueller matrices are alternative representations of the same physical phenomenology, $\mathbf{M}_J$ is normal, nonnormal, or degenerate, inasmuch its associated Jones matrix is.

Lu and Chipman defined the *nonnormality parameter* $\eta$ (called by them the *inhomogeneity parameter*) [22] as

$$\eta = |\cos(\theta/2)| \quad (22)$$

where $\theta$ is the angle subtended between the Poincaré vectors of the eigenstates. $\eta$ reaches its minimum value, $\eta = 0$, for normal Mueller matrices, while $\eta = 1$ corresponds to media whose pure eigenstates are coincident (degenerate $\mathbf{M}_J$).

The expression of $\eta$ provided by Lu and Chipman in terms of the Jones matrix $\mathbf{T}$, can be translated to the following form in terms of the corresponding Mueller matrix $\mathbf{M}_J$

$$\eta^2 = \frac{4m_{00} - |T|^2 - \left|T^2 - 4\sqrt[4]{\det\mathbf{M}_J}\right|}{4m_{00} - |T|^2 + \left|T^2 - 4\sqrt[4]{\det\mathbf{M}_J}\right|} \quad (23.a)$$

with

$$T \equiv \mathrm{tr}\,\mathbf{T} = \frac{1}{\sqrt{2}}\frac{\mathrm{tr}\,\mathbf{M}_J + m_{01} + m_{10} + i(m_{32} + m_{23})}{\sqrt{m_{00} + m_{11} + m_{01} + m_{10}}} \quad (23.b)$$

The *normality* of $\mathbf{M}_J$ can be measured by means of the parameter $\mu \equiv |\sin(\theta/2)|$ given by

$$\mu^2 \equiv 1 - \eta^2 = \frac{2\left|T^2 - 4\sqrt[4]{\det\mathbf{M}_J}\right|}{4m_{00} - |T|^2 + \left|T^2 - 4\sqrt[4]{\det\mathbf{M}_J}\right|} \quad (24)$$

Let us now recall that any pure $\mathbf{M}_J$ can be expressed as

$$\mathbf{M}_J = \mathbf{M}_{R2}\,\mathbf{M}_{DL0}(m_{00}, D)\,\mathbf{M}_{R1} \quad (25)$$

where $\mathbf{M}_{DL0}(m_{00}, D)$ is the Mueller matrix of a horizontal linear diattenuator, while $\mathbf{M}_{R1}$ and $\mathbf{M}_{R2}$ represent respective retarders (in general elliptic). Normality is achieved when $\mathbf{M}_{R2} = \mathbf{M}_{R1}^T$, and thus a general form of a normal $\mathbf{M}_J$ is

$$\mathbf{M}_{RD} = \mathbf{M}_R^T\,\mathbf{M}_{DL0}(m_{00}, D)\,\mathbf{M}_R \quad (26)$$

showing that a normal pure Mueller matrix depends on five real parameters, namely $m_{00}$, $D$, the retardance $\Delta$ of $\mathbf{M}_R$ and the azimuth $\varphi$ and ellipticity angle $\chi$ of the fast eigenstate of $\mathbf{M}_R$.

## 5. Two-vector representation of a pure Mueller matrix

From the polar decomposition in Eq. (**6**) of a pure Mueller matrix, it turns out that $\hat{\mathbf{M}}_J$ is completely characterized by $\mathbf{D}$ and $\bar{\mathbf{R}}$, or alternatively by $\mathbf{P}$ and $\bar{\mathbf{R}}$. Let us analyze below the different possible cases depending on the value of the diattenuation-polarizance $D$ of $\hat{\mathbf{M}}_J$ and on the relative orientation of $\bar{\mathbf{R}}$ and $\mathbf{D}$.

### 5.1. Neutral filter

A nondepolarizing medium exhibiting zero polarizance-diattenuation ($D = 0$) and zero retardance ($\Delta = 0$) is represented by a Mueller matrix $\mathbf{M}_J = m_{00}\mathbf{I}$ proportional to the identity matrix $\mathbf{I}$, and therefore its two-vector representation reduces to two zero vectors $\mathbf{D} = \bar{\mathbf{R}} = \mathbf{0}$, represented by the origin of the Poincaré sphere. See Table 1.

### 5.2. Retarder

When the nondepolarizing medium does not exhibit diattenuation ($D = 0$), then its Mueller matrix has necessarily the form

$$\mathbf{M}_J = m_{00}\mathbf{M}_R = m_{00}\begin{pmatrix} 1 & \mathbf{0}^T \\ \mathbf{0} & \mathbf{m}_R \end{pmatrix} \quad (27)$$

so that $\hat{\mathbf{M}}_J$ is the Mueller matrix of a retarder, and hence it is fully determined by the corresponding representative vector $\bar{\mathbf{R}}$, to which the two-vector representation is reduced (vector $\mathbf{D} = \mathbf{0}$ is represented by the origin of the Poincaré sphere). Obviously, since $\mathbf{M}_R$ is an orthogonal matrix, it is normal ($\eta = 0$). See Table 1.

### 5.3. Normal diattenuator

When $\Delta = 0$, that is, $\bar{\mathbf{R}} = \mathbf{0}$, or equivalently $\mathbf{M}_R = \mathbf{I}$, and $0 < D < 1$, the nondepolarizing medium does not exhibit retardance and behaves like a normal partial diattenuator ($\mu = 1$) whose associated symmetric Mueller matrix $\mathbf{M}_D$ has the form shown in Eq. (**9**). In this case, $\mathbf{P} = \mathbf{D}$, and the two-vector representation reduces to vector $\mathbf{D}$ ($\bar{\mathbf{R}}$ is represented by the origin of the Poincaré sphere).

A peculiarity of nondepolarizing media with $0 < D < 1$ is that, unlike for totally polarized input states, its action reduces the degree of polarization of some partially polarized input states [24], which is a consequence of the entangled nature of diattenuation and depolarization [13]. See Table 1.

### 5.4. Normal polarizer

When $\Delta = 0$ (i.e., $\bar{\mathbf{R}} = \mathbf{0}$) and $D = 1$, then $\mathbf{M}_J$ is symmetric and singular, so that necessarily $\mathbf{P} = \mathbf{D}$ ($\mu = 1$). The general form of the Mueller matrix of a normal polar polarizer is

$$\mathbf{M}_{\hat{D}} = m_{00} \begin{pmatrix} 1 & \hat{\mathbf{D}}^T \\ \hat{\mathbf{D}} & \hat{\mathbf{D}} \otimes \hat{\mathbf{D}}^T \end{pmatrix} \quad \left(|\hat{\mathbf{D}}| \equiv D = 1\right) \tag{28}$$

The two vector representation of $\mathbf{M}_{\hat{D}}$ reduces to the vector $\hat{\mathbf{D}}$, whose end point touches the surface of the unit sphere, while $\bar{\mathbf{R}} = \mathbf{0}$, which is represented by the origin of the unit sphere. See Table 1.

Nondepolarizing media satisfying $D = 1$ are characterized by the fact that the output light (for both forward and reverse interactions) is always fully polarized, regardless of the degree of polarization of the input light.

### 5.5. Nonnormal diattenuator

When $\Delta \neq 0$ and $0 < D < 1$, the medium exhibits both retardance and diattenuation. If, in addition, $\bar{\mathbf{R}}$ and $\mathbf{D}$ are nonparallel, the medium is called a *nonnormal diattenuator*. In this case, diattenuation and polarizance vectors are necessarily different, $\mathbf{P} \neq \mathbf{D}$. As with normal diattenuators, the action of a nonnormal diattenuator reduces the degree of polarization of some partially polarized input states. In this case, $\mathbf{M}_J$ is nonsingular ($\det \mathbf{M}_J \neq 0$) and nonnormal ($\mu < 1$). The two vector representation of a nonnormal diattenuator is given by both nonzero vectors $\bar{\mathbf{R}}$ and $\mathbf{D}$ (see Table 1).

The polarizance vector is given by $\mathbf{P} = \mathbf{m}_R \mathbf{D}$, that represents a rotation of $\mathbf{D}$ by an angle (retardance)

$$\Delta = \pi |\bar{\mathbf{R}}| = \arccos \left\{ \left[ \mathrm{tr}\left(\mathbf{M}_J \mathbf{M}_D^{-1}\right) \right]/2 - 1 \right\} \tag{29}$$

about the *retardance axis* defined by the direction of $\bar{\mathbf{R}}$.

### 5.6. Diattenuating retarder

When $\Delta \neq 0$ and $0 < D < 1$, the medium exhibits both retardance and diattenuation. If, in addition, $\bar{\mathbf{R}}$ and $\mathbf{D}$ are parallel, then diattenuation and polarizance vectors are necessarily equal, $\mathbf{P} = \mathbf{D}$, and the medium is called *diattenuating retarder* [19], or *nonideal retarder* [16], and can be considered a serial combination of a normal diattenuator and a retarder with coincident eigenstates, so that the Mueller matrix is normal ($\mu = 1$) and has the form shown in Eq. (**26**), or, from the polar decomposition (**10**) of $\mathbf{M}_{RD}$, it turns out that $\mathbf{M}_{RD}$ can always be expressed as

$$\mathbf{M}_{RD} = m_{00} \begin{pmatrix} 1 & \mathbf{D}^T \\ \mathbf{D} & \mathbf{m}_{RD} \end{pmatrix}$$

$$\mathbf{m}_{RD} \equiv \sin\kappa \, \mathbf{m}_R + \frac{1-\sin\kappa}{\cos^2\kappa}\left(\mathbf{D}\otimes\mathbf{D}^T\right) \tag{30}$$

Therefore, diattenuating retarders are characterized by a two-vector representation where $\bar{\mathbf{R}}$ and $\mathbf{D}$ are nonzero and parallel, with $D < 1$. See Table 1.

### 5.7. Nonnormal polarizer

When $D = 1$, the medium behaves as a polarizer, whose Mueller matrix can always be expressed as [8]

$$\mathbf{M}_{\hat{P}\hat{D}} = m_{00} \begin{pmatrix} 1 & \hat{\mathbf{D}}^T \\ \hat{\mathbf{P}} & \hat{\mathbf{P}} \otimes \hat{\mathbf{D}}^T \end{pmatrix} \quad \left(D \equiv |\hat{\mathbf{D}}| = |\hat{\mathbf{P}}| = 1\right) \tag{31}$$

If, in addition to $D = 1$, $\bar{\mathbf{R}}$ and $\mathbf{D}$ are not parallel (i.e. $\hat{\mathbf{P}} \neq \hat{\mathbf{D}}$), then $\mathbf{M}_{\hat{P}\hat{D}}$ is not symmetric (hence nonnormal, $\mu < 1$) and the medium is called a *nonnormal polarizer*. Note that, in this case, the pair ($\mathbf{D}$, $\mathbf{P}$) also determines completely the Mueller matrix $\mathbf{M}_{\hat{P}\hat{D}}$. The equivalent retarder is not uniquely determined, but by applying the criterion of minimum retardance [8,22], whose appropriateness is physically well founded and justified [25], the direction of $\bar{\mathbf{R}}$ is that of the vector product $\mathbf{D} \wedge \mathbf{P}$, while the associated minimum retardance is given by

$$\Delta_m = \arccos\left(\mathbf{P}^T \mathbf{D}\right) \tag{32}$$

or, in terms of $\mu$, [22]

$$\Delta_m = 2\arccos\mu \tag{33}$$

Particularly interesting is the special case of a *degenerate polarizer*, characterized by the combined conditions $D = 1$ and $\Delta = \pi$, that is, $\mathbf{P} = -\mathbf{D}$, with associated Mueller matrix

$$\mathbf{M}_{\hat{D}\pm} = m_{00} \begin{pmatrix} 1 & \hat{\mathbf{D}}^T \\ -\hat{\mathbf{D}} & -\hat{\mathbf{D}}\otimes\hat{\mathbf{D}}^T \end{pmatrix} \quad \left(D \equiv |\hat{\mathbf{D}}| = 1\right) \tag{34}$$

## 6. Plane representation of a pure Mueller matrix

Unlike the two-vector representation of $\mathbf{M}_J$ by means of the pair of three-dimensional vectors $\bar{\mathbf{R}}$ and $\mathbf{D}$, the dual linear retarder decomposition of $\mathbf{M}_J$ constitutes the basis for a simple geometric representation of $\mathbf{M}_J$ through the set of two-dimensional vectors composed of the diattenuation and straight retardance vectors of the central linear diattenuating retarder $\mathbf{M}_{RDL0}$ ($\mathbf{D}_0$ and $\bar{\mathbf{R}}_0$ respectively) together with the straight retardance vectors $\bar{\mathbf{R}}_{L1}$ and $\bar{\mathbf{R}}_{L2}$ of the entrance and exit linear retarders $\mathbf{M}_{RL1}$ and $\mathbf{M}_{RL2}$ respectively

$$\mathbf{D}_0 \equiv (D, 0, 0)^T \quad \bar{\mathbf{R}}_0 \equiv (\Delta_0/\pi, 0, 0)^T$$
$$\bar{\mathbf{R}}_{L1} \equiv \bar{\mathbf{R}}(\mathbf{M}_{RL1}) \quad \bar{\mathbf{R}}_{L2} \equiv \bar{\mathbf{R}}(\mathbf{M}_{RL2}) \tag{35}$$

These vectors can be represented in the equatorial section of the Poincaré sphere as shown in Figure 1.

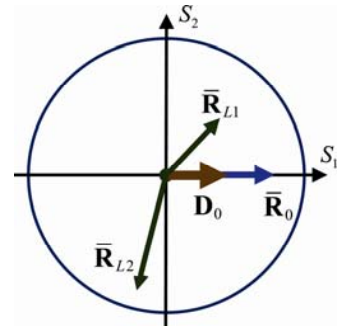

**Figure 1**. Plane representation of a pure Mueller matrix $\mathbf{M}_J$ by means of two points on the $S_1$ axis representing diattenuation and intrinsic retardance respectively, together with the linear retardance vectors of the entrance and exit linear retarders of the dual linear retarder decomposition of $\mathbf{M}_J$.

$\mathbf{D}_0$ and $\bar{\mathbf{R}}_0$ are represented by respective points on the $S_1$ axis, while $\bar{\mathbf{R}}_{L1}$ and $\bar{\mathbf{R}}_{L2}$ are two-dimensional vectors in the plane $S_1 S_2$. The particular features of this kind of representation for the different types of pure Mueller matrices are briefly analyzed in the next subsections.

### 6.1. Neutral filter

In this case, $\mathbf{M}_J$ is proportional to the identity matrix and therefore $\mathbf{D}_0 = \bar{\mathbf{R}}_0 = \bar{\mathbf{R}}_{L1} = \bar{\mathbf{R}}_{L2} = \mathbf{0}$. See Table 2.

### 6.2. Retarder

In the case of a retarder, the dual linear retarder decomposition (**15**) of its Mueller matrix $\mathbf{M}_R$ can always be performed as follows, with the choice $\Delta_0 = 0$,

$$\mathbf{M}_R = \mathbf{M}_{RL2} \mathbf{M}_{RL1} \tag{36}$$

where $\mathbf{M}_{RL1}$ and $\mathbf{M}_{RL2}$ are determined from $\mathbf{M}_R$ in not a unique way (because both linear retarders involve a total of four parameters versus the three free parameters of a general retarder [16]). The plane representation of $\mathbf{M}_R$ reduces to the pair $(\bar{\mathbf{R}}_{L1}, \bar{\mathbf{R}}_{L2})$ (see Table 2). The expressions for the parameters of $\mathbf{M}_R$ in terms of those of $\mathbf{M}_{RL1}$ and $\mathbf{M}_{RL2}$ can be found in Ref. [26].

The ambiguity in the plane representation of $\mathbf{M}_R$ is a peculiarity of retarders, which, in turn, are straightforwardly characterized by the fact that $\mathbf{M}_R$ is an orthogonal Mueller matrix, whose general expression is given by Eq. (**8**). Therefore, for the simple case of retarders, the plane representation is not advantageous in comparison to the two-vector representation. Nevertheless, the said ambiguity does not reduce the interest of the plane decomposition for all the remaining cases, in which $D > 0$.

### 6.3. Normal diattenuator

In this case $\mathbf{M}_{RL2} = \mathbf{M}_{RL1}^T$ (i.e., $\bar{\mathbf{R}}_{L2} = -\bar{\mathbf{R}}_{L1}$) and, since $\text{tr} \mathbf{M}_D = 2(1 + \sin \kappa)$, then $\Delta_0 = 0$. Therefore, the plane representation of the Mueller matrix $\mathbf{M}_D$ of the normal diattenuator is given by $\mathbf{D}_0$ (point $D$ on the $S_1$ axis), and the pair of vectors $(\bar{\mathbf{R}}_{L1}, -\bar{\mathbf{R}}_{L1})$. See Table 2.

### 6.4. Normal polarizer

This case is analogous to that of a normal diattenuator, but with the peculiarity that $D = 1$, so that $\mathbf{D}_0$ touches the unit circle at point $(1,0)$. See Table 2.

### 6.5. Nonnormal diattenuator

In this generic case (excluding normal Mueller matrices and polarizers), $\bar{\mathbf{R}}_{L2} \neq -\bar{\mathbf{R}}_{L1}$ and the four vectors $(\mathbf{D}_0, \bar{\mathbf{R}}_0, \bar{\mathbf{R}}_{L1}, \bar{\mathbf{R}}_{L2})$ are required for a complete characterization of $\hat{\mathbf{M}}_J$. See Table 2.

### 6.6. Diattenuating retarder

For a diattenuating retarder $\mathbf{M}_{RD}$ (whose eigenstates are, in general, elliptic), $\Delta_0 \neq 0$, $0 < D < 1$, and $\bar{\mathbf{R}}_{L2} = -\bar{\mathbf{R}}_{L1}$. The corresponding plane representation of $\mathbf{M}_{RD}$ is shown in Table 2.

### 6.7. Nonnormal polarizer

This case corresponds to $D = 1$, and $\bar{\mathbf{R}}_{L2} \neq -\bar{\mathbf{R}}_{L1}$, while $\Delta_0$ is undetermined and can be taken as $\Delta_0 = 0$ without loss of generality. The features of plane representation of a nonnormal polarizer are shown in Table 2.

In the special case of a degenerate polarizer, from Eqs. (**21**) and (**34**), it turns out that the following relation must be satisfied

$$\mathbf{m}_{RL1} \mathbf{m}_{RL2} = \begin{pmatrix} -1 & 0 & 0 \\ 0 & \cos\alpha & \sin\alpha \\ 0 & \sin\alpha & -\cos\alpha \end{pmatrix} \tag{37}$$

and consequently the product $\mathbf{M}_{RL1} \mathbf{M}_{RL2}$ is the Mueller matrix of an elliptic retarder with azimuth $\pi/4$, retardance $\pi$, and arbitrary ellipticity. In this case, the absolute values and azimuths of $\bar{\mathbf{R}}_{L1}$ and $\bar{\mathbf{R}}_{L2}$ are mutually linked in a complicated manner, and therefore the identification of degenerate polarizers in the plane representation is not as straightforward as in the two-vector representation.

## 7. Conclusion

From the polar decomposition of a pure (or nondepolarizing) Mueller matrix $\mathbf{M}_J \equiv m_{00} \hat{\mathbf{M}}_J$ it turns out that $\mathbf{M}_J$ is determined by the scalar quantity $m_{00}$ together the pair of constitutive vectors $\bar{\mathbf{R}}$ and $\mathbf{D}$ (or alternatively by $\bar{\mathbf{R}}$ and $\mathbf{P}$). The magnitudes and orientations of these vectors in the Poincaré sphere determine completely the properties of $\hat{\mathbf{M}}_J$. Through the two-vector representation, the polarimetric behavior of the nondepolarizing medium is easily identified as pertaining to one of the following possible cases (see Table 1)

- $\bar{\mathbf{R}} = \mathbf{D} = \mathbf{0}$. Neutral filter, $\mathbf{M}_J = m_{00} \mathbf{I}$.
- $\bar{\mathbf{R}} \neq \mathbf{0}$, $\mathbf{D} = \mathbf{0}$. Retarder, $\mathbf{M}_J = m_{00} \mathbf{M}_R$, Eq. (**8**).
- $\bar{\mathbf{R}} = \mathbf{0}$, $0 < |\mathbf{D}| < 1$. Normal diattenuator, $\mathbf{M}_J = \mathbf{M}_D$ Eq. (**9**).
- $\bar{\mathbf{R}} = \mathbf{0}$, $|\mathbf{D}| = 1$. Normal polarizer, $\mathbf{M}_J = \mathbf{M}_{\hat{D}}$ Eq. (**28**).
- $\bar{\mathbf{R}} \neq \mathbf{0}$, $0 < |\mathbf{D}| < 1$, with $\bar{\mathbf{R}}$ and $\mathbf{D}$ nonparallel ($\mathbf{P} \neq \mathbf{D}$). Nonnormal diattenuator, Eq. (**10**).
- $\bar{\mathbf{R}} \neq \mathbf{0}$, $0 < |\mathbf{D}| < 1$, with $\bar{\mathbf{R}} \parallel \mathbf{D}$ ($\mathbf{P} = \mathbf{D}$). Diattenuating retarder, $\mathbf{M}_J = \mathbf{M}_{RD}$ Eq. (**26**).
- $0 < |\bar{\mathbf{R}}| < 1$, $|\mathbf{D}| = 1$, with $\bar{\mathbf{R}}$ and $\mathbf{D}$ being nonparallel. Nonnormal polarizer $\mathbf{M}_J = \mathbf{M}_{\hat{P}\hat{D}}$, Eq. (**31**).

Moreover, from the dual linear retarder decomposition of a pure Mueller matrix $\mathbf{M}_J$, a *plane representation* of $\hat{\mathbf{M}}_J$ has been described in terms of vectors $\mathbf{D}_0$ and $\bar{\mathbf{R}}_0$ (both lying in the $S_1$ axis of the Poincaré sphere), together with the pair of vectors $\bar{\mathbf{R}}_{L1}$ and $\bar{\mathbf{R}}_{L2}$ (both lying in the equatorial plane $S_1 S_2$ of the Poincaré sphere). The magnitudes and orientations of these vectors determine completely the properties of $\hat{\mathbf{M}}_J$. The polarimetric behavior of the nondepolarizing medium is easily identified as pertaining to one of the following possible cases (see Table 2):

- $\mathbf{D}_0 = \bar{\mathbf{R}}_0 = \bar{\mathbf{R}}_{L1} = \bar{\mathbf{R}}_{L2} = \mathbf{0}$. Neutral filter, $\mathbf{M}_J = m_{00} \mathbf{I}$.
- $\mathbf{D}_0 = \mathbf{0}$. Retarder, $\mathbf{M}_J = m_{00} \mathbf{M}_R$, Eq. (**8**).
- $\bar{\mathbf{R}}_0 = \mathbf{0}$, $0 < |\mathbf{D}_0| < 1$, $\bar{\mathbf{R}}_{L2} = -\bar{\mathbf{R}}_{L1}$. Normal diattenuator, $\mathbf{M}_J = \mathbf{M}_D$ Eq. (**9**).
- $\bar{\mathbf{R}}_0 = \mathbf{0}$, $|\mathbf{D}_0| = 1$, $\bar{\mathbf{R}}_{L2} = -\bar{\mathbf{R}}_{L1}$. Normal polarizer, $\mathbf{M}_J = \mathbf{M}_{\hat{D}}$ Eq. (**28**).
- $\bar{\mathbf{R}}_0 \neq \mathbf{0}$, $0 < |\mathbf{D}| < 1$, $\bar{\mathbf{R}}_{L2} \neq -\bar{\mathbf{R}}_{L1}$. Nonnormal diattenuator, Eq. (**10**).
- $\bar{\mathbf{R}}_0 \neq \mathbf{0}$, $0 < |\mathbf{D}| < 1$, $\bar{\mathbf{R}}_{L2} = -\bar{\mathbf{R}}_{L1}$. Diattenuating retarder, $\mathbf{M}_J = \mathbf{M}_{RD}$ Eq. (**26**).
- $\bar{\mathbf{R}}_0 = \mathbf{0}$, $|\mathbf{D}_0| = 1$, $\bar{\mathbf{R}}_{L2} \neq -\bar{\mathbf{R}}_{L1}$. Nonnormal polarizer $\mathbf{M}_J = \mathbf{M}_{\hat{P}\hat{D}}$, Eq. (**31**).

In summary, the two-vector representation features a geometric view of a normalized pure Mueller matrix in terms of the pair of three-dimensional vectors $\mathbf{D}$ and $\bar{\mathbf{R}}$, while the plane representation is based on a set of two points along the axis $S_1$, given by the end points of vectors $\mathbf{D}_0$ and $\bar{\mathbf{R}}_0$, together with the pair of straight linear retardance vectors $\bar{\mathbf{R}}_{L1}$ and $\bar{\mathbf{R}}_{L2}$ lying in the equatorial plane $S_1 S_2$ of the Poincaré sphere. While the two-vector representation has, in general, a more direct interpretation, the plane representation is constituted by vectors lying in a common plane. Moreover, while the two-vector representation is always unambiguous, the plane representation requires establishing certain conventions for the representation of the simple case of pure retarders.

Examples of the two-vector and plane representations of some of the pure Mueller matrices with nonzero diattenuation that are commonly encountered in experimental measurements (after the corresponding filtering process, where appropriate) are presented in Table 3.

**Funding Information.** Ministerio de Economía y Competitividad (FIS2011-22496 and FIS2014-58303-P); Gobierno de Aragón (E99).

**Table 1. Two-vector representation of a pure Mueller matrix (1)**

| | Retardance and polarizance | Retardance and diattenuation vectors | Normality | Mueller matrix | Representative vectors |
|---|---|---|---|---|---|
| Neutral filter | $\Delta = 0$<br>$D = 0$ | $\bar{\mathbf{R}} = \mathbf{0}$<br>$\mathbf{D} = \mathbf{0}$ | $\mu = 1$ | $\mathbf{M}_J = m_{00}\mathbf{I}$ | |
| Retarder | $\Delta \neq 0$<br>$D = 0$ | $\bar{\mathbf{R}} \neq \mathbf{0}$<br>$\mathbf{D} = \mathbf{0}$ | $\mu = 1$ | $\mathbf{M}_R = m_{00}\begin{pmatrix} 1 & \mathbf{0}^T \\ \mathbf{0} & \mathbf{m}_R \end{pmatrix}$ | |
| Normal diattenuator | $\Delta = 0$<br>$0 < D < 1$ | $\bar{\mathbf{R}} = \mathbf{0}$<br>$\mathbf{D} \neq \mathbf{0}$ | $\mu = 1$ | $\mathbf{M}_D \equiv m_{00}\begin{pmatrix} 1 & \mathbf{D}^T \\ \mathbf{D} & \mathbf{m}_D \end{pmatrix}$ | |
| Normal polarizer | $\Delta = 0$<br>$D = 1$ | $\bar{\mathbf{R}} = \mathbf{0}$<br>$|\mathbf{D}| = 1$ | $\mu = 1$ | $\mathbf{M}_{\hat{D}} = m_{00}\begin{pmatrix} 1 & \hat{\mathbf{D}}^T \\ \hat{\mathbf{D}} & \hat{\mathbf{D}}\otimes\hat{\mathbf{D}}^T \end{pmatrix}$ | |
| Nonnormal diattenuator | $\Delta \neq 0$<br>$0 < D < 1$ | $\bar{\mathbf{R}} \neq \mathbf{0}$<br>$\mathbf{D} \neq \mathbf{0}$<br>$\dfrac{\mathbf{D}}{|\mathbf{D}|} \neq \dfrac{\bar{\mathbf{R}}}{|\bar{\mathbf{R}}|}$ | $\mu < 1$ | $\mathbf{M}_J = m_{00}\begin{pmatrix} 1 & \mathbf{D}^T \\ \mathbf{m}_R\mathbf{D} & \mathbf{m}_R\mathbf{m}_D \end{pmatrix}$ | |

**Table 1. Two-vector representation of a pure Mueller matrix (2)**

| | Retardance and polarizance | Retardance and diattenuation vectors | Normality | Mueller matrix | Representative vectors |
|---|---|---|---|---|---|
| Diattenuating retarder | $\Delta \neq 0$<br>$0 < D < 1$ | $\bar{\mathbf{R}} \neq \mathbf{0}$<br>$\mathbf{D} \neq \mathbf{0}$<br>$\dfrac{\mathbf{D}}{|\mathbf{D}|} = \dfrac{\bar{\mathbf{R}}}{|\bar{\mathbf{R}}|}$ | $\mu = 1$ | $\mathbf{M}_{RD} = m_{00} \begin{pmatrix} 1 & \mathbf{D}^T \\ \mathbf{D} & \mathbf{m}_R \mathbf{m}_D \end{pmatrix}$ | |
| Nonnormal polarizer | $\Delta \neq 0$<br>$D = 1$ | $\bar{\mathbf{R}} \neq \mathbf{0}$<br>$|\mathbf{D}| = 1$ | $\mu < 1$ | $\mathbf{M}_{\hat{P}\hat{D}} = m_{00} \begin{pmatrix} 1 & \hat{\mathbf{D}}^T \\ \hat{\mathbf{P}} & \hat{\mathbf{P}} \otimes \hat{\mathbf{D}}^T \end{pmatrix}$ | |
| Degenerate polarizer | $\Delta = \pi$<br>$D = 1$ | $|\bar{\mathbf{R}}| = 1$<br>$|\mathbf{D}| = 1$ | $\mu = 0$ | $\mathbf{M}_{\hat{D}\pm} = m_{00} \begin{pmatrix} 1 & \hat{\mathbf{D}}^T \\ -\hat{\mathbf{D}} & -\hat{\mathbf{D}} \otimes \hat{\mathbf{D}}^T \end{pmatrix}$ | |

**Table 2. Plane representation of a pure Mueller matrix (1)**

| | Polarizance and intrinsic retardance | Representative vectors | Normality | Mueller matrix | Representative vectors |
|---|---|---|---|---|---|
| Neutral filter | $D=0$ <br> $\Delta_0 = 0$ | $\mathbf{D}_0 = \mathbf{0}$ <br> $\bar{\mathbf{R}}_0 = \mathbf{0}$ <br> $\bar{\mathbf{R}}_{L1} = \mathbf{0}$ <br> $\bar{\mathbf{R}}_{L2} = \mathbf{0}$ | $\mu = 1$ | $\mathbf{M}_J = m_{00}\mathbf{I}$ | |
| Retarder | $D=0$ <br> $\Delta_0 \neq 0$ | $\mathbf{D}_0 = \mathbf{0}$ <br> $\bar{\mathbf{R}}_0 = \mathbf{0}$ <br> $\bar{\mathbf{R}}_{L1} \neq \mathbf{0}$ <br> $\bar{\mathbf{R}}_{L2} \neq \mathbf{0}$ | $\mu = 1$ | $\mathbf{M}_R = m_{00}\begin{pmatrix} 1 & \mathbf{0}^T \\ \mathbf{0} & \mathbf{m}_R \end{pmatrix}$ | |
| Normal diattenuator | $0 < D < 1$ <br> $\Delta_0 = 0$ | $\mathbf{D}_0 \neq \mathbf{0}$ <br> $\bar{\mathbf{R}}_0 = \mathbf{0}$ <br> $\bar{\mathbf{R}}_{L1} = -\bar{\mathbf{R}}_{L2}$ | $\mu = 1$ | $\mathbf{M}_D \equiv m_{00}\begin{pmatrix} 1 & (\mathbf{m}_{RL}\mathbf{D}_0)^T \\ \mathbf{m}_{RL}\mathbf{D}_0 & \mathbf{m}_{RL}^T\mathbf{m}_{DL0}\mathbf{m}_{RL} \end{pmatrix}$ | |
| Normal polarizer | $D=1$ <br> $\Delta_0 = 0$ | $|\mathbf{D}_0| = 1$ <br> $\bar{\mathbf{R}}_0 = \mathbf{0}$ <br> $\bar{\mathbf{R}}_{L1} = -\bar{\mathbf{R}}_{L2}$ | $\mu = 1$ | $\mathbf{M}_D \equiv m_{00}\begin{pmatrix} 1 & (\mathbf{m}_{RL}\hat{\mathbf{D}}_0)^T \\ \mathbf{m}_{RL}\hat{\mathbf{D}}_0 & \mathbf{m}_{RL}\mathbf{m}_{\hat{D}L0}\mathbf{m}_{RL}^T \end{pmatrix}$ | |
| Nonnormal diattenuator | $0 < D < 1$ <br> $\Delta_0 \neq 0$ | $\mathbf{D}_0 \neq \mathbf{0}$ <br> $\bar{\mathbf{R}}_0 \neq \mathbf{0}$ <br> $\bar{\mathbf{R}}_{L1} \neq -\bar{\mathbf{R}}_{L2}$ | $\mu < 1$ | $\mathbf{M}_J \equiv m_{00}\begin{pmatrix} 1 & (\mathbf{m}_{RL1}^T\mathbf{D}_0)^T \\ \mathbf{m}_{RL2}\mathbf{D}_0 & \mathbf{m}_{RL2}\mathbf{m}_{RDL0}\mathbf{m}_{RL1} \end{pmatrix}$ | |
| Diattenuating retarder | $0 < D < 1$ <br> $\Delta_0 \neq 0$ | $\mathbf{D}_0 \neq \mathbf{0}$ <br> $\bar{\mathbf{R}}_0 \neq \mathbf{0}$ <br> $\bar{\mathbf{R}}_{L1} = -\bar{\mathbf{R}}_{L2}$ | $\mu = 1$ | $\mathbf{M}_{RD} \equiv m_{00}\begin{pmatrix} 1 & (\mathbf{m}_{RL}\mathbf{D}_0)^T \\ \mathbf{m}_{RL}\mathbf{D}_0 & \mathbf{m}_{RL}\mathbf{m}_{RDL0}\mathbf{m}_{RL}^T \end{pmatrix}$ | |

**Table 2. Plane representation of a pure Mueller matrix (2)**

| | Polarizance and intrinsic retardance | Representative vectors | Normality | Mueller matrix | Representative vectors |
|---|---|---|---|---|---|
| Nonnormal polarizer | $D = 1$ <br> $\Delta_0 = 0$ | $\mathbf{D}_0 \neq \mathbf{0}$ <br> $\overline{\mathbf{R}}_0 = \mathbf{0}$ <br> $\overline{\mathbf{R}}_{L1} \neq -\overline{\mathbf{R}}_{L2}$ | $\mu < 1$ | $\mathbf{M}_{\hat{P}\hat{D}} = m_{00} \begin{pmatrix} 1 & \hat{\mathbf{D}}^T \\ \hat{\mathbf{P}} & \hat{\mathbf{P}} \otimes \hat{\mathbf{D}}^T \end{pmatrix}$ <br><br> $\hat{\mathbf{D}} = \mathbf{m}_{RL1}^T \mathbf{D}_0 \quad \hat{\mathbf{P}} = \mathbf{m}_{RL2} \mathbf{D}_0$ | (figure: Poincaré plane with axes $S_1$, $S_2$ showing vectors $\overline{\mathbf{R}}_{L1}$, $\overline{\mathbf{R}}_{L2}$, $\mathbf{D}_0$ on unit circle) |

**Table 3. Two-vector and plane representations of common types of pure Mueller matrices**

| Type | Mueller matrix | Two-vector representation | Plane representation |
|---|---|---|---|
| Horizontal linear diattenuating retarder | $\begin{pmatrix} 1 & \cos\kappa & 0 & 0 \\ \cos\kappa & 1 & 0 & 0 \\ 0 & 0 & \sin\kappa\cos\Delta_0 & \sin\kappa\sin\Delta_0 \\ 0 & 0 & -\sin\kappa\sin\Delta_0 & \sin\kappa\cos\Delta_0 \end{pmatrix}$ | | |
| Horizontal linear diattenuator | $m_{00}\begin{pmatrix} 1 & \cos\kappa & 0 & 0 \\ \cos\kappa & 1 & 0 & 0 \\ 0 & 0 & \sin\kappa & 0 \\ 0 & 0 & 0 & \sin\kappa \end{pmatrix}$ | | |
| Horizontal linear polarizer | $m_{00}\begin{pmatrix} 1 & 1 & 0 & 0 \\ 1 & 1 & 0 & 0 \\ 0 & 0 & 0 & 0 \\ 0 & 0 & 0 & 0 \end{pmatrix}$ | | |
| Circular diattenuator | $m_{00}\begin{pmatrix} 1 & 0 & 0 & \cos\kappa \\ 0 & \sin\kappa & 0 & 0 \\ 0 & 0 & \sin\kappa & 0 \\ \cos\kappa & 0 & 0 & 1 \end{pmatrix}$ | | |
| Circular polarizer | $m_{00}\begin{pmatrix} 1 & 0 & 0 & 1 \\ 0 & 0 & 0 & 0 \\ 0 & 0 & 0 & 0 \\ 1 & 0 & 0 & 1 \end{pmatrix}$ | | |